\documentstyle[epsf]{mn}

\newif\ifAMStwofonts

\def\etal{{et al. \rm}}
\def\x{X1908+075}
\def\xs{X1908+075 }
\def\brs{Br$\gamma$ }

\ifoldfss
  \ifCUPmtlplainloaded \else
    \NewTextAlphabet{textbfit} {cmbxti10} {}
    \NewTextAlphabet{textbfss} {cmssbx10} {}
    \NewMathAlphabet{mathbfit} {cmbxti10} {} 
    \NewMathAlphabet{mathbfss} {cmssbx10} {} 
  \fi
  \ifAMStwofonts
    \ifCUPmtlplainloaded \else
      \NewSymbolFont{upmath} {eurm10}
      \NewSymbolFont{AMSa} {msam10}
      \NewMathSymbol{\upi}     {0}{upmath}{19}
      \NewMathSymbol{\umu}     {0}{upmath}{16}
      \NewMathSymbol{\upartial}{0}{upmath}{40}
      \NewMathSymbol{\leqslant}{3}{AMSa}{36}
      \NewMathSymbol{\geqslant}{3}{AMSa}{3E}

    \fi
  \fi
\fi 

\ifnfssone
  \newmathalphabet{\mathit}
  \addtoversion{normal}{\mathit}{cmr}{m}{it}
  \addtoversion{bold}{\mathit}{cmr}{bx}{it}
  \newmathalphabet{\mathbfit} 
  \addtoversion{normal}{\mathbfit}{cmr}{bx}{it}
  \addtoversion{bold}{\mathbfit}{cmr}{bx}{it}
  \newmathalphabet{\mathbfss} 
  \addtoversion{normal}{\mathbfss}{cmss}{bx}{n}
  \addtoversion{bold}{\mathbfss}{cmss}{bx}{n}
  \ifAMStwofonts
    \ifCUPmtlplainloaded \else
      \UseAMStwoboldmath
      \makeatletter
      \new@mathgroup\upmath@group
      \define@mathgroup\mv@normal\upmath@group{eur}{m}{n}
      \define@mathgroup\mv@bold\upmath@group{eur}{b}{n}
      \edef\UPM{\hexnumber\upmath@group}
      \new@mathgroup\amsa@group
      \define@mathgroup\mv@normal\amsa@group{msa}{m}{n}
      \define@mathgroup\mv@bold\amsa@group{msa}{m}{n}
      \edef\AMSa{\hexnumber\amsa@group}
      \makeatother
      \mathchardef\upi="0\UPM19
      \mathchardef\umu="0\UPM16
      \mathchardef\upartial="0\UPM40
      \mathchardef\leqslant="3\AMSa36
      \mathchardef\geqslant="3\AMSa3E
    \fi
  \fi
\fi 

\ifnfsstwo
  \DeclareMathAlphabet{\mathbfit}{OT1}{cmr}{bx}{it}
  \SetMathAlphabet\mathbfit{bold}{OT1}{cmr}{bx}{it}
  \DeclareMathAlphabet{\mathbfss}{OT1}{cmss}{bx}{n}
  \SetMathAlphabet\mathbfss{bold}{OT1}{cmss}{bx}{n}
  \ifAMStwofonts
    \ifCUPmtlplainloaded \else
      \DeclareSymbolFont{UPM}{U}{eur}{m}{n}
      \SetSymbolFont{UPM}{bold}{U}{eur}{b}{n}
      \DeclareSymbolFont{AMSa}{U}{msa}{m}{n}
      \DeclareMathSymbol{\upi}{0}{UPM}{"19}
      \DeclareMathSymbol{\umu}{0}{UPM}{"16}
      \DeclareMathSymbol{\upartial}{0}{UPM}{"40}
      \DeclareMathSymbol{\leqslant}{3}{AMSa}{"36}
      \DeclareMathSymbol{\geqslant}{3}{AMSa}{"3E}
    \fi
  \fi
\fi 

\ifCUPmtlplainloaded \else
  \ifAMStwofonts \else 
    \def\upi{\pi}
    \def\umu{\mu}
    \def\upartial{\partial}
  \fi
\fi

\title[A new OB-supergiant X-ray binary]{Near-infrared identification of the counterpart to \x: a new OB-supergiant X-ray binary}
\author[T. Morel and Y. Grosdidier]
{\parbox{179mm}{\begin{flushleft}
\vspace{-0.5cm}
{\LARGE T. Morel$^{1}$} 
\thanks{e-mail: morel@astropa.unipa.it}
{\LARGE and Y. Grosdidier$^{2,3}$}
\end{flushleft}
}\vspace*{0.200cm}\\  
\parbox{159mm}{
$^1$ Istituto Nazionale di Astrofisica, Osservatorio Astronomico di Palermo G. S. Vaiana, Piazza del Parlamento 1, I-90134 Palermo, Italy\\
$^2$ Instituto de Astrof\'{\i}sica de Canarias, Calle V\'{\i}a L\'actea s/n, E-38200 La Laguna (Tenerife), Spain\\
$^3$ Department of Physics, McGill University, 3600 University St., Montr\'eal, Qu\'ebec, Canada, H3A 2T8
}}

\date{Accepted ???.
      Received ???;
      in original form ??? }

\pagerange{\pageref{firstpage}--\pageref{lastpage}}
\pubyear{2001}

\begin{document}

\maketitle

\label{firstpage}

\begin{abstract}
We report the near-infrared (IR) identification of the likely counterpart to \x, a  highly-absorbed Galactic X-ray source recently suspected to belong to the rare class of OB supergiant-neutron star binary systems. Our {\it JHK$_s$}-band imaging of the field reveals the existence within the X-ray error boxes of a near-IR source consistent with an early-type star lying at $d$$\sim$7 kpc and suffering $A_V$$\sim$16 mag of extinction, the latter value being in good agreement with the hydrogen column density derived from a modelling of the X-ray spectrum. Our follow-up, near-IR spectroscopic observations confirm the nature of this candidate and lead to a late O-type supergiant classification, thereby supporting the identification of a new Galactic OB-supergiant X-ray binary. 
\end{abstract}


\section{Introduction}
The class of High-Mass X-Ray Binaries (HMXRBs) is defined as systems made up of a compact object (generally a neutron star) accreting material from the wind of an early-type companion, either an OB supergiant or a Be star (in which case X-ray outbursts occur when the neutron star passes through the circumstellar disk). Although the vast majority of the HMXRBs are identified as Be/X-ray binaries, the few systems with an OB supergiant primary are of particular interest. First, these objects might ultimately lead to the formation of a neutron star-black hole system and are thus important in the context of massive binary star evolution. Second, the disturbance induced by the orbital motion of the collapsed object through the primary stellar outflow can be used to probe the structure and physical properties of radiation-driven winds (e.g., Haberl, White \& Kallman 1989; Kaper, Hammerschlag-Hensberge \& van Loon 1993). To date, however, only a handful of OB-supergiant X-ray binary systems have been optically identified (Liu, van Paradijs \& van den Heuvel 2000). Therefore, there is a need to uncover other members of this important population.

The Galactic X-ray source \xs was first detected using the {\it Uhuru} satellite (Forman \etal 1978) and has been persistently catalogued ever since by several X-ray missions. These observations are summarized by Wen, Remillard \& Bradt (2000; hereafter WRB) who present a variability study of this source using the All-Sky Monitor (ASM) onboard the {\it Rossi X-Ray Timing Explorer (RXTE)} satellite. The X-ray power ($<$$L_X$$>$$\sim$5 $\times$ 10$^{36}$ ergs s$^{-1}$ in the 5--100 keV energy band for an adopted distance of 7 kpc and $A_V$$\sim$15 mag) showed a coherent, energy-dependent sinusoidal modulation ($\pm$30 per cent in the 5--12 keV energy band) over 3 yrs with a period of 4.400$\pm$0.001 days. This value is best interpreted as the orbital period of a binary system, but seems too long to be associated with a low-mass X-ray binary. The X-ray luminosity is also typical of other HMXRBs, but the smooth and regular nature of the X-ray light curve argues against a transient Be-neutron star binary system. These arguments suggest that the mass-donor star is an OB-type supergiant, while the hard X-ray spectrum supports a neutron star companion. Levine \etal (2004) recently discussed pointed observations of \xs carried out with the Proportional Counter Array (PCA) and the High Energy X-ray Timing Experiment (HEXTE) onboard {\it RXTE}. Their analysis confirmed the modulation of the X-ray flux according to the 4.400-d orbital period, but also revealed the existence of pulsed emission with a period of 605 s. The location of \xs in the pulse period/orbital period plane clearly indicates that it is a wind-fed accretion system (Corbet 1986). The mass function derived from the Doppler delay curve, along with an estimate of the orbital inclination angle derived from a modelling of the orbital phase-dependence of the X-ray flux ($38^o \la i \la 72^o$), led to a mass of the primary in the range 9--31 M$_{\odot}$ and an upper limit on its size of about 22 R$_{\odot}$. After assumptions about the outflow dynamics, the wind column density estimates were used to derive a wind mass-loss rate, $\dot{M}$, exceeding 1.3 $\times$ 10$^{-6}$ M$_{\odot}$ yr$^{-1}$. Based on the inferred stellar parameters and high mass-loss rate, they further hypothesized that the mass donor could be a Wolf-Rayet star. 

The {\it Einstein} imaging proportional counter (IPC) has localized this X-ray source at 19$^{\rm h}$10$^{\rm m}$46$^{\rm s}$ and +07$^{\circ}$36$'$07$''$ (J2000) with an uncertainty of about 50 arcsec. Additional constraints on the position are provided by observations made with the modulation collimator experiment (A-3) onboard the {\it HEAO 1} satellite (see WRB).\footnote{Note that the position of the grid of {\it HEAO 1} A-3 ``diamonds'' in fig.1 of WRB is incorrect (L. W. Wen 2001, private communication). See Fig.1 for the correct position.} This relatively small error box offers promising prospects for seeking the counterpart, although the very high dust obscuration resulting both from the presumably large distance and from the location close to the Galactic plane ($b$$\sim$--0.8$^{\circ}$) prevents any follow-up studies in the optical. Here we present near-IR imaging and spectroscopic observations identifying the likely counterpart to this X-ray source as a late O-type supergiant.

\section{Observations and results}
\subsection{Imaging}
We obtained {\it JHK$_s$}-band images of the field on service observing mode on August 9, 2001 (JD 2,452,131.50) with the IR camera CAIN-II mounted on the Carlo S\'anchez Telescope (TCS) at Teide Observatory (Canary Islands, Spain). The observations were carried out under photometric sky conditions and a seeing of about 1.6 arcsec. In the adopted configuration, the plate scale is 1 arcsec pixel$^{-1}$ and the field of view is 4.2 $\times$ 4.2 arcmin$^2$. 

Standard reduction procedures for near-IR imaging were applied (multiplication by a bad pixel mask, bias subtraction, removal of pixel-to-pixel sensitivity variations in the array). A mean sky frame was created by applying a median filtering procedure to the data frames themselves. After subtraction of this mean sky frame, the individual science images (which were obtained in a dither sequence with a spatial offset of 64 arcsec) were aligned and mosaicked. The astrometric calibration was achieved with reference to the positions of 16 field stars in the USNO-A2.0 catalogue. The internal astrometric errors are of the order of 0.35 arcsec, as judged from the residuals between the measured and catalogued positions of the reference stars. The near-IR images, along with an optical image taken from the Digitized Sky Survey, are shown in Fig.1. Our images reach a limiting magnitudes in all three bands comparable to the 2MASS data, but our resolution is considerably higher owing to the poor weather conditions experienced during the 2MASS observations (seeing $\sim$2.6 arcsec). 

The stellar magnitudes were computed for the stars detected at the 3$\sigma$ level using the task {\tt phot} in the {\tt IRAF}\footnote{{\tt IRAF} is distributed by the National Optical Astronomy Observatories, operated by the Association of Universities for Research in Astronomy, Inc., under cooperative agreement with the National Science Foundation.} photometric package {\tt apphot}. The stellar magnitudes were determined by integrating the counts through a circular aperture with a fixed radius of 3 pixels. This value provides a robust estimate of the stellar flux, while minimizing the contribution of background noise. The local sky level was estimated in a nearby, concentric annular region. The faint standard star FS35 was used for calibration (Hunt \etal 1998).\footnote{We transformed the {\it K} magnitude quoted by Hunt \etal (1998) into a {\it K$_s$} magnitude using equation (8) of Carpenter (2001). Although this implicitely assumes that the CAIN and 2MASS {\it K$_s$}-band total responses are similar, the errors arising from such an approximation are very small for a star with the colour properties of FS35 (see, e.g., equations (16) and (26) of Carpenter 2001).} A subsequent correction for atmospheric extinction was applied. The limiting magnitudes for a 3$\sigma$ detection in the science frames are about 17.8, 16.5 and 16.5 mag in {\it J}, {\it H} and {\it K$_s$}, respectively. The counterpart is expected to be detected in all three bands if we consider an early-type star with the distance and amount of interstellar absorption inferred from the X-ray data. 

The colour-colour and colour-magnitude diagrammes for the stars in common between the {\it Einstein} and {\it HEAO 1} error boxes are shown in Fig.2 (we also plot using different symbols the three brightest sources to the South-East of the field lying within the {\it HEAO 1} error box, but just outside the {\it Einstein} error circle). To isolate the potential counterparts, we have indicated the expected positions of two representative early-type stars (O5 V and O9 I) lying at $d$=7$\pm$3 kpc and suffering $A_V$=15$\pm$5 mag of extinction (WRB). These rather generous uncertainties are somewhat arbitrary and are only intended at leading to a conservative pre-selection of the counterparts. Three candidates (hereafter A, B and C; see also Fig.1 for their position) appear to have near-IR properties consistent with the assumed distance and amount of visual extinction. Candidate C, however, formally lies outside the intersection of the two X-ray error boxes. The coordinates and near-IR magnitudes of these three potential counterparts are quoted in Table 1.

\begin{figure*}
\epsfbox{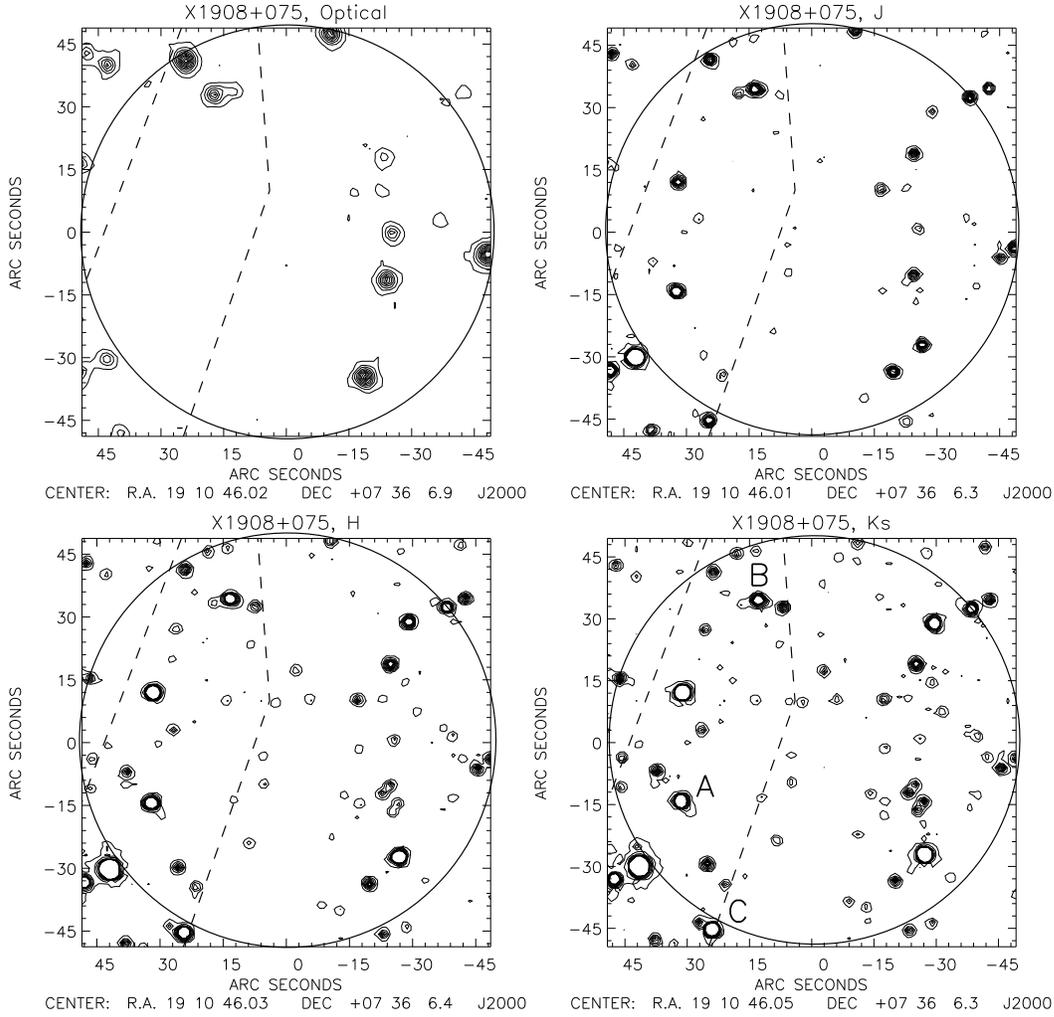}
\caption{Contour maps of the field in the optical (image taken from a POSS-II red plate), {\it J}, {\it H} and {\it K$_s$} bands (the lowest contour is drawn at 3$\sigma$ above the background noise, and then by steps of 6$\sigma$). The 50-arcsec error circle of the {\it Einstein} satellite ({\it solid line}), as well as the diamond-like error box of the {\it HEAO 1} satellite ({\it dashed line}) are overlayed. The IDs of the potential candidates (A, B and C, see text) are indicated in the lower, right-hand panel. North is up and East is to the left.}
\end{figure*}

\begin{figure*}
\epsfbox{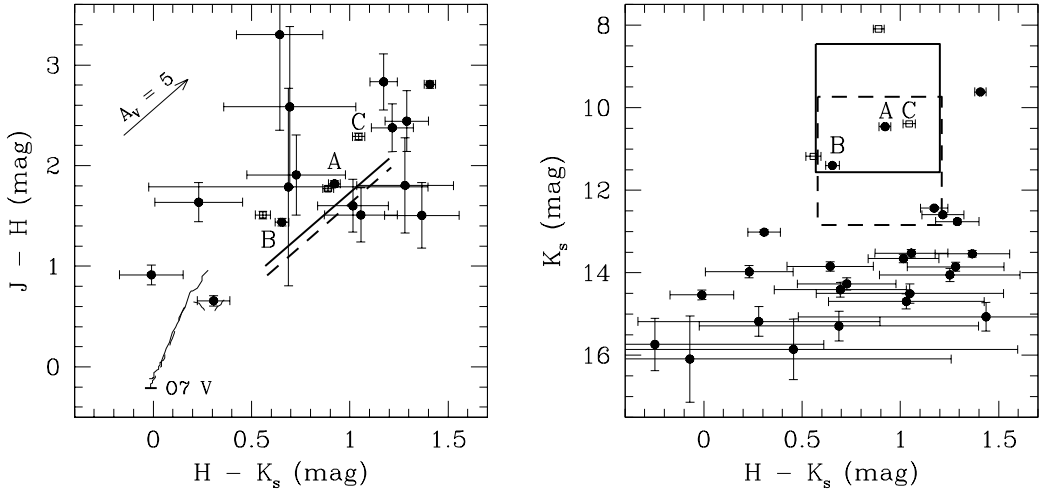}
\caption{Colour-colour and colour-magnitude diagrammes for the stars in common between the {\it Einstein} and {\it HEAO 1} error boxes ({\it filled circles}). Only stars detected in the three bands are shown. The three brightest sources to the South-East of the field lying within the {\it HEAO 1} error box, but just outside the {\it Einstein} error circle are also plotted ({\it open circles}). The position of candidates A, B and C is indicated. The diagonal lines in the left-hand panel and the boxes in the right-hand panel show the expected loci of the near-IR counterpart to \x, assuming an O5 V ({\it dashed line}) or an O9 I ({\it solid line}) star (see text). We used the calibrated $M_V$ values and near-IR colours of Vacca \etal (1996) and Koornneef (1983), respectively. The intrinsic colours of dwarfs ({\it dashed line}) and supergiants ({\it solid line}) in the 2MASS photometric system are shown in the left-hand panel (we used the data of Koornneef 1983 and the transformation equations of Carpenter 2001). An arrow indicates the effect of reddening (near-IR interstellar extinction law from Rieke \& Lebofsky 1985).}
\end{figure*}

\begin{table*}
\caption{Coordinates and near-IR magnitudes of the three potential counterparts.}
\begin{tabular}{cccccc}
\hline
ID  & $\alpha$ (J2000)$^a$ & $\delta$ (J2000)$^a$          &  {\it J} (mag)$^b$ & {\it H} (mag)$^b$ & {\it K$_s$} (mag)$^b$\\\hline
A   & 19h 10m 48.204s (21) & +07$^{\circ}$ 35$'$ 52.32$''$ (26)  &  13.199$\pm$0.018  & 11.380$\pm$0.012  &  10.457$\pm$0.018 \\
B   & 19h 10m 46.923s (18) & +07$^{\circ}$ 36$'$ 41.13$''$ (30)  &  13.492$\pm$0.021  & 12.054$\pm$0.014  &  11.399$\pm$0.021 \\
C   & 19h 10m 47.713s (28) & +07$^{\circ}$ 35$'$ 20.91$''$ (26)  &  13.721$\pm$0.025  & 11.432$\pm$0.012  &  10.387$\pm$0.018 \\\hline 
\end{tabular}\\
\begin{flushleft}
$^a$ The numbers given in parentheses are the 1-$\sigma$ uncertainties in the last decimal places. They take into account the typical astrometric uncertainty of the USNO-A2.0 catalogue ($\sim$0.25 arcsec), as well as differences in the relative position of the source in the {\it J}, {\it H} and {\it K$_s$} frames. The coordinates of candidate A differ by $\sim$0.7 arcsec from the values quoted in the 2MASS Point Source Catalogue.\\
$^b$ Note that these values may be slightly affected by nearby, faint near-IR sources (see Fig.1). The quoted errors are the 1-$\sigma$ uncertainties. 
\end{flushleft}
\end{table*}

\subsection{Spectroscopy}
Follow-up, medium-resolution {\it HK}-band spectroscopy of the two prime candidates A and B was obtained on service observing mode on May 24, 2002 with the near-IR spectrograph CGS4 mounted on the United Kingdom Infrared Telescope (UKIRT). The sky was photometric and the seeing in the range 0.4--0.6 arcsec. The observations were carried out at JD 2,452,419.04, which corresponds to an orbital phase $\phi$$\sim$0.07 according to the ephemeris of Levine \etal (2004) for a circular orbit: $\cal{P}$=4.4006$\pm$0.0006 days and $T_{\rm 90}$=2,452,643.8 (zero phase is fixed at minimum X-ray flux, i.e., at superior conjunction of the neutron star). We used the 300-mm focal length camera, a 0.6-arcsec wide slit and the 40 lines mm$^{-1}$ grating in the first order (resolving power $R$$\sim$800 at 2 $\umu$m). The spectral range covered in a single exposure is 1.60--2.24 $\umu$m. The slit of the spectrograph was oriented East-West, except for candidate B (PA=--15$^{\circ}$, i.e. nearly North-South) in order to avoid contamination from a faint near-IR source immediately lying to the East (see Fig.1). Several background-limited exposures (for a total integration time amounting to 150 and 300 s for candidates A and B, respectively) were obtained while nodding along the slit. 

Initial reduction tasks (application of the bad pixel mask, dark and bias subtraction, as well as flat-fielding) were carried out with the CGS4 data reduction pipeline ORAC-DR. Further reduction steps were carried out with the {\tt IRAF} software. The OH sky lines in the science frames were removed by subtracting two-dimensional spectra obtained at different slit positions. The spectra were optimally extracted taking into account the data values and detector characteristics. An argon lamp taken at the beginning of the observations was used for the wavelength calibration. The spectra were subsequently continuum-normalized by fitting a low-order cubic spline polynomial to (pseudo) line-free regions. The removal of the telluric lines was performed using spectra of the A3 star HD 179939 obtained close in time (less than 50 min) and airmass (less than 0.03) to the target observations. The spectrum of the A star can be considered to first order featureless, except for the presence of the hydrogen lines of the Brackett series (Br 10--12 and \brs in the {\it H} and {\it K} band, respectively). The procedure for removing the atmospheric telluric lines in the science spectra consisted of: (a) interactively fitting the hydrogen lines in the telluric standard using Gaussian or Voigt profiles, paying special attention not to include the weak  telluric features on both sides of the Brackett lines, (b) then subtracting the fitted profiles from the A-star spectrum, and (c) finally dividing out the target spectra by this template telluric spectrum. The typical signal-to-noise ratio estimated from the photon statistics is 130 and 60 at \brs for candidates A and B, respectively. These figures are likely upper limits, however, considering the significant source of error arising from the unavoidably imperfect telluric correction. The {\it HK}-band spectra of the two candidates are shown in Fig.3.

\begin{figure*}
\epsfbox{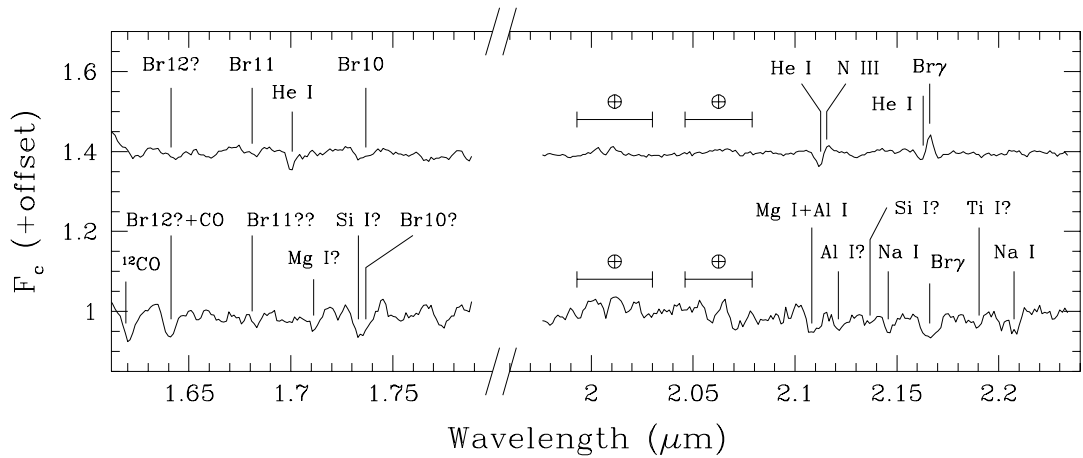}
\caption{Continuum-normalized {\it HK}-band spectra of candidates A ({\it top}) and B ({\it bottom}). The spectrum of candidate A has been shifted along the ordinate axis by 0.4 continuum units. The most prominent spectral features are labelled (uncertain identifications are indicated by a question mark). The line wavelengths and EWs of the lines identified in candidate A are quoted in Table 2. The reader is referred to Lan\c con \& Rocca-Volmerange (1992), Meyer \etal (1998) and Wallace \& Hinkle (1996, 1997) for further details regarding the features identified in candidate B. The $\oplus$ symbols mark the position of the strong CO$_2$ telluric lines. Some spurious features arising from an imperfect telluric subtraction can be seen at these locations.}
\end{figure*}

Although some of the weak spectral features identified in candidate B might be spurious in view of the rather limited data quality, this star clearly exhibits a large number of metallic lines absent either in candidate A or in the spectrum of a G5 star (HD 179220) taken during the same observing run (not shown). The identification of the spectral lines was mainly based on the atlases of Meyer \etal (1998) and Wallace \& Hinkle (1996, 1997).  The presence of the luminosity-sensitive, second-overtone $^{12}$CO(6,3) band head at 1.619 $\umu$m can be taken as evidence for a cool star ($T_{\rm eff}$ $\la$ 5000 K), most likely of luminosity class I-III (Meyer \etal 1998; Ivanov \etal 2004). The marginal detection of candidate B in the optical waveband suggests that it is a foreground source (Fig.1). 

The spectral morphology of candidate A is drastically different, with only evidence for hydrogen, helium and nitrogen lines. Several pieces of evidence unambiguously point to an early-type star (Hanson \& Conti 1994; Hanson, Conti \& Rieke 1996; Hanson, Rieke \& Luhman 1998; Hanson \etal 2004, in prep.; Lenorzer \etal 2004): (a) the emission-like N\,III $\lambda$2.1155 $\umu$m line is only seen in O stars and, in particular, in members of the 'Of' class, defined as showing the optical N\,III $\lambda$$\lambda$4634, 4640 \AA \ doublet in emission (Conti \& Alschuler 1971); (b) \brs is only in emission in O supergiant stars, {\it not} in O dwarfs/giants or B-type stars; (c) the absence of the He\,II lines at 1.693 and 2.189 $\umu$m, as well as the C\,IV lines at 2.069, 2.078 and 2.083 $\umu$m, rules out an early O-type star. This assertion is supported by the detection of He\,I $\lambda$2.113 $\umu$m. The measured strength of several diagnostic lines (see Table 2) all converge to a classification as a late O-type star, while \brs in emission diagnoses a supergiant (Hanson \etal 1996, 1998). Based on these arguments, we propose a O7.5--O9.5 If classification. The {\it K}-band spectral morphology is reminiscent of HD 57060 (O7 Iafp) or HD 151804 (O8 Iaf; Hanson \etal 1996). Similarities are also found with HD 123008 (ON9.5 Iab), for instance (Hanson \etal 2004, in prep.). The most conspicuous difference is the absence in \xs of the He\,I $\lambda$2.058 $\umu$m line, as observed in other OB-supergiant X-ray binaries (e.g., Cyg X-1; Hanson \etal 1996). Setting stringent limits on the strength on this feature is hampered by the existence of a strong telluric band stretching from 2.045 to 2.080 $\umu$m. If present, however, the helium line is clearly very weak. We note the lack of a clear relationship in O stars between the strength of this feature and the spectral type (Hanson \etal 1996). This peculiar behaviour may result from the sensitivity of this line to the extreme-UV flux, for instance (e.g., Najarro \etal 1994). 

The question naturally arising is to what extent a classification scheme based on 'normal' OB stars can be confidently used for stars harbouring a compact object. Despite the fact that Hanson \etal (1996) obtained {\it K}-band spectra of several HMXRBs and did not find clear evidence for spectral peculiarities compared to single stars with the same optical classification, we cannot certainly rule out that the presence of the neutron star affects in some way our spectral classification. However, the fact that all diagnostic lines yield consistent results gives us confidence in our results. Furthermore, a circular orbit would imply that our spectroscopic observations were secured close to superior conjunction of the neutron star (see Sect. 2.2). Would it be the case, this fortunate circumstance would imply that the spectral type we derive is not strongly biased by the X-ray irradiated part of the primary's photosphere.

\begin{table}
\caption{Line properties of \x.}
\begin{tabular}{lc}
\hline
Line & EW (\AA)\\\hline
Br 12 $\lambda$1.6412  & +1.4\\
Br 11 $\lambda$1.6810  & +0.8\\
He\,I $\lambda$1.7007  & +1.9\\
Br 10 $\lambda$1.7367  & +1.5\\
He\,I $\lambda$2.1127  & +1.2\\
N\,III $\lambda$2.1155  & --0.6\\
He\,I $\lambda$2.163  & +0.7\\
\brs $\lambda$2.1661  & --1.4\\\hline
\end{tabular}\\
\end{table}

\section{Discussion}
Although phase-locked photometric variations might be expected in \x, our near-IR magnitudes appear to be compatible with the 2MASS Point Source Catalogue data (observations obtained at JD 2,451,394.61; designation 2MASS19104821+0735516): $J$=13.228$\pm$0.021, $H$=11.457$\pm$0.027 and $K_s$=10.480$\pm$0.022 mag. Using the absolute magnitudes appropriate for an O9 I star (Vacca, Garmany \& Shull 1996; Koornneef 1983) and the near-IR interstellar extinction law of Rieke \& Lebofsky (1985), we obtain from our data: $d$$\sim$7 kpc and $A_V$$\sim$16.5 mag. WRB and Levine \etal (2004) derived a column density of intervening interstellar material to the source in the range $\cal N_{\rm H}$=3.0--4.6 $\times$ 10$^{22}$ atoms cm$^{-2}$ from a modelling of the X-ray spectrum. This translates into an amount of interstellar extinction, $A_V$=16.0--24.6 mag (Bohlin, Savage \& Drake 1978). This range of values is in reasonable agreement with our estimate considering the large uncertainties involved, for instance, in the empirical transformation between the hydrogen column density and the optical extinction in the {\it V} band. \xs is undetected on a POSS-II red plate going down to $R$$\sim$20 mag (see Fig.1), in accordance with our derived spectral type, distance and amount of interstellar extinction (we expect $R$$\sim$24.5 mag for $A_V$$\sim$16.5 mag, $M_V$=--6.5 and $V$--$R$=--0.15; Vacca \etal 1996).

Levine \etal (2004) estimated the following physical parameters for the primary star based on {\it RXTE} data: $M_{\star}$=9--31 M$_{\odot}$, $R_{\star}$ $\la$ 22 R$_{\odot}$ and $\dot{M}$ $\ga$ 1.3 $\times$ 10$^{-6}$ M$_{\odot}$ yr$^{-1}$. The constraint on the stellar radius, together with the high mass-loss rate inferred, led them to propose that the massive component could be a Wolf-Rayet star. This interpretation is, however, not supported by the lack in \xs of strong, broad emission lines (see, e.g., Morris \etal 1996; Figer, McLean \& Najarro 1997). Because Wolf-Rayet stars have absolute magnitudes and colours comparable to O-type supergiants, it is likely that such an object would have been isolated by its photometric properties (Fig.2). Although perhaps not compelling in itself, systems made up of a Wolf-Rayet star and a compact companion are expected to have short orbital periods following a spiral-in phase (e.g., de Donder, Vanbeveren \& van Bever 1997). A period of 4.8 hr is observed in Cygnus X--3, the prime candidate for such a system (van Kerkwijk \etal 1996). 

The most stringent constraint coming from the X-ray data is the fact that the neutron star is accreting material well below the Eddington rate. This implies that the companion is wind-fed and lies within the Roche lobe, i.e., $R_{\star}$ $\la$ 22 R$_{\odot}$. Optical radial velocity curves have been combined with X-ray eclipse data by van Kerkwijk, van Paradijs \& Zuiderwijk (1995) to derive accurate masses and radii for three B0--B0.5 supergiants in X-ray binaries. They derived $R_{\star}$$\sim$30 R$_{\odot}$ and $M_{\star}$$\sim$24 M$_{\odot}$ for Vela X-1, while $R_{\star}$$\sim$15 R$_{\odot}$ and $M_{\star}$$\sim$15 M$_{\odot}$ was found for 4U1538--52 and SMC X-1. This shows that a supergiant classification does not enter in conflict with the upper limit on the stellar radius, although it is likely that \xs is close to filling its Roche lobe, as commonly observed in these systems. 

Mass-loss rates of late O-type supergiants measured from the radio free-free emission or from the strength of the H$\alpha$ line typically lie in the range 1--10 $\times$ 10$^{-6}$ M$_{\odot}$ yr$^{-1}$ (e.g., Leitherer 1988; Lamers \& Leitherer 1993). Theoretical models of radiation-driven winds predict an {\it inverse} dependence of the mass-loss rate with the stellar mass (e.g., Abbott 1982). The value $\dot{M}$ $\ga$ 1.3 $\times$ 10$^{-6}$ M$_{\odot}$ yr$^{-1}$ derived by Levine \etal (2004) is thus entirely conceivable considering that \xs has a lower mass that the stars discussed above. However, a comparable luminosity is required for this object: typically log\,($L_{\star}$/$L_{\odot}$) $\ga$ 5.2 (Vink, de Koter \& Lamers 2000). We note that the N\,III $\lambda$2.116 line in emission suggests an evolved O star with a very high luminosity and a rather strong stellar wind (Conti \& Alschuler 1971).  

\section{Conclusion}
The use of several diagnostic lines in the {\it HK} windows has led to a robust classification of the likely counterpart to \xs as a late O-type star, while \brs in emission strongly points to a supergiant (Hanson \etal 1996). The proposed O7.5--O9.5 If spectroscopic classification is internally consistent with the near-IR photometric properties and also compatible with the stellar parameters independently derived by Levine \etal (2004) from X-ray data, provided that the star satisfies log\,($L_{\star}$/$L_{\odot}$) $\ga$ 5.2. As mentioned previously, the X-ray data do not argue in favour of a Be/X-ray binary system. The lack of strong near-IR emission lines in \xs strongly supports this conclusion (Everall \etal 1993). Despite the prohibitively high visual extinction, further advances in our understanding of this system may come from radio or time-resolved near-IR observations, the latter both in the photometric and spectroscopic modes. Further X-ray observations with higher spatial resolution, as well as near-IR spectroscopic data for other possible candidates in the field (notably candidate C; see Figs 1 and 2) are also needed for an unambiguous identification of the counterpart.

\section*{Acknowledgments}
We acknowledge A. M. Levine for constructive and fruitful discussions during the course of this work. We also wish to thank the anonymous referee for useful comments and M. M. Hanson for making her manuscript available to us prior to publication. The imaging data used in this paper were obtained as part of the Teide Observatory Service Observing Programme. The United Kingdom Infrared Telescope is operated by the Joint Astronomy Centre on behalf of the U.K. Particle Physics and Astronomy Research Council. The spectroscopic data reported here were obtained as part of the UKIRT Service Programme. The Digitized Sky Survey was produced at the Space Telescope Science Institute (STScI) under US government grant NAGW-2166. The images of these surveys are based on photographic data obtained using the Oschin-Schmidt Telescope on Palomar Mountain and the UK Schmidt Telescope. The plates were processed into the present compressed digital form with the permission of these institutions. This publication makes use of data products from the Two Micron All Sky Survey, which is a joint project of the University of Massachusetts and the Infrared Processing and Analysis
Center/California Institute of Technology, funded by the National Aeronautics and Space Administration and the National Science Foundation.

\bsp

\label{lastpage}

\end{document}